# Towards Democratized Academic LLM Hosting

## Practical insights from needs, implementations, and governance across Dutch institutions

Azza E Ahmed[1], Sreeparna Deb[1], Taylor Stone[1], Sven Warris[2], Aaron Ding[1], Joost Bengsch[1], Margriet Cox[1], Julio Alexandrino de Oliveira Filho[3], Alexandre Strube[4], Avishek Anand[1], Thomas Abeel[1], and the "Towards Democratized Academic LLM Hosting" Workshop Consortium*

[1] Delft University of Technology; [2] Rabobank; [3] TNO; [4] Juelich Supercomputing Centre

## Abstract

Academic LLM hosting is emerging as an operational infrastructure domain with recurring socio-technical patterns, tensions, and governance requirements across institutions. Based on a workshop with academic infrastructure, research-support, education, and policy-adjacent stakeholders, this paper synthesizes current practices and open questions around institutionally governed LLM services for Dutch academia. It highlights challenges around deployment models, energy and utilization, governance, tooling, commercial integration, and research and education use cases, and identifies emerging directions for service pathways, coordination, and responsible operation.

## 1 Introduction

Large Language Model (LLM) capabilities are now actively needed across academic institutions, supporting education, research, guidance and support, learning analytics, and a myriad of other institutional workflows. This need can be facilitated by using local hardware and open-source tooling/models, while retaining institutional autonomy over data, models, and service provision. However, the needed hardware is expensive and limited. Equally, the engineering effort can be quite extensive for delivering robust, secure and compliant LLM endpoints.

This creates a broader institutional challenge. Academic LLM hosting is not simply a choice between self-hosting and commercial APIs, but a governance and infrastructure problem of routing different academic use cases through a portfolio of local, shared, and commercial service pathways.

To examine these questions in practice, we organized the workshop "Towards Democratized Academic LLM Hosting" (https://www.4tu.nl/nirict/Projects/All_projects/Towards%20Democratized%20Academic%20LLM%20Hosting/ ) on 16 March 2026 at Mondai | House of AI, Delft, NL. The workshop brought together over 80 participants from Dutch academic institutions, national infrastructure organizations, European public-interest generative AI (genAI) hosting initiatives, and related research-support communities, including infrastructure providers, research software engineers, data stewards, educators, researchers, and policy/governance support roles. It was designed as a practice-oriented discussion rather than a policy exercise, with keynotes, breakout sessions, and a panel focusing on the operational realities of providing LLM capabilities in academic settings.

The conversation focused on practical challenges and strategic decisions around deploying, managing, and scaling LLM services in research, education and institutional contexts. Central themes included model selection, infrastructure constraints, user guidance, governance, energy and utilization, commercial integration, and the balance between local, shared, and commercial genAI offerings.

This paper synthesizes the workshop discussions across the keynotes, breakout sessions, and closing panel. It should therefore be read as a set of practice-derived observations and interpretations, not as statistically representative findings.

The contribution of this paper is threefold: 1) it frames academic LLM hosting as a socio-technical infrastructure problem; 2) it identifies recurring operational tensions across Dutch academic institutions; and 3) it proposes a portfolio/service-pathway perspective for matching use cases to local, shared, and commercial infrastructure options.

The main questions are: who would actually use such a platform - students, researchers, educators, professional-services staff, or all university users? How is such infrastructure currently approached within academia today: what is the state-of-affairs, what are the emerging trade-offs, and which questions remain open?

The remainder of this paper first outlines the institutional context for academic LLM hosting, then synthesizes the workshop discussions across the main thematic areas, and finally identifies recurring tensions, emerging directions, and open questions.

# 2 Why this matters now: The case for LLM hosting

## 2.1 Institutional drivers and growing demand

Institutional demand for LLM access in academia has rapidly expanded across research, education, and administrative contexts. This demand is no longer limited to individual experimentation or small technical user groups; it is increasingly visible at the scale of courses, departments, and whole institutions.

The demand is also diverse. Non-technical departments use LLMs heavily for translation, summarization, and related text-processing tasks. Researchers use them for tasks such as analyzing interview transcripts, managing metadata, extracting information from documents, and translating research needs into prompts or downstream workflows, in addition to coding and software assistance. Academic publishers and journals are also exploring the use of generative AI in editorial and production workflows (Hanneke et al. 2026; Cohen and Moher 2025), further broadening institutional demand. In education, even a single course can require substantial and sustained access: for example, a 300-student AI-augmented engineering course at TU Delft running over 10 weeks with weekly assignments requires stable, high-quality, and fair API access throughout the teaching period. Students also use LLMs informally for everyday study tasks, including text processing, coding, and other forms of academic support outside formally designed assignments. This diversity of use cases makes it difficult to treat LLM access as a narrow technical service for specialist users only.

At institutional scale, universities are negotiating data-safe agreements, piloting multiple genAI tools, and purchasing bulk licences for commercial LLM services. Microsoft Copilot is already part of this landscape across Dutch institutions, although adoption models vary, and privacy and confidentiality concerns continue to shape how access is provided (“Privacy Risks Microsoft 365 Copilot Remain ‘Orange’ despite Improvements | SURF.Nl” 2026). In parallel, network-level observations at some institutions suggest that large volumes of unmanaged usage already exist: during the workshop, one institution reported around 3.5 million API calls per month from its network to commercial genAI providers. Similar extrapolations in other research contexts suggest that fully relying on commercial APIs for research-related genAI needs could become financially significant at institutional scale, while also exposing institutions to changing token-based pricing models and creating tensions with digital sovereignty goals.

Equitable access is another institutional driver. In education, differences already arise not only between students who can afford premium commercial genAI tools and those who rely on free or open-source alternatives, but also in how accessible different tools are to students with different needs and abilities. This raises a practical educational concern: unequal access to high-quality genAI tools may affect grades, learning outcomes, and participation.

Across LLM-hosting initiatives, data privacy remains one of the clearest drivers for local or institutionally governed hosting. Commercial API use can require privacy officers to negotiate contracts, assess data-processing conditions, and monitor compliance, which was repeatedly raised in the workshop as an operational burden for institutions. This burden grows when each new tool or provider requires separate review. In some cases, a clear privacy boundary around external data sharing becomes the direct trigger for moving away from unmanaged commercial APIs toward self-hosted or institutionally controlled infrastructure.

Other drivers relate to digital sovereignty, institutional control, and long-term academic reliability. These include control over data, uncertainty about how large commercial providers handle institutional or sensitive information, use of public funding for overseas commercial services, control over models and guardrails, and intellectual property considerations, and security concerns around export controls and dual-use technologies.

These local pressures also sit within a broader compute-access problem: Dutch science has reported growing demand for research computing capacity, while AI research more broadly has become increasingly shaped by unequal access to large-scale compute resources (“Onderzoek” 2025; Ahmed and Wahed 2020).

Finally, several discussions pointed to domain-specific models as a potentially stronger institutional case than trying to compete with large generalist models from major technology companies. The argument for academic LLM hosting is therefore not that universities should replicate frontier commercial systems. Rather, the case is strongest where institutions need stable, governed, equitable, and domain-relevant LLM capabilities that align with academic research, education, and public-sector responsibilities.

### 2.2 From experimentation to institutional/operational service provision

The surge of genAI tooling and providers has resulted in numerous standards, practices and integration options. For local, data-conscious, and privacy-conscious deployments, the endless supply of tooling can itself be overwhelming.

Many academic institutions still have limited capacity to run LLM capabilities as reliable services. At the same time, the level of commercial-provider use is not always well understood institutionally, making it harder to plan local services, governance, and support capacity.

GenAI is treated differently across institutional contexts: in some cases primarily as a productivity tool, and in others as the subject of research itself, requiring experimentation, customization, and domain-specific adaptation.

For research, model persistence is also important: when commercial models are deprecated, changed, or withdrawn during a project, experiments may need to be repeated and results become harder to reproduce.

These pressures shift LLM access from an individual tool choice toward an institutional service question, as illustrated in Figure 1. Production readiness requires much more than exposing model inference end points. Robust deployments need orchestration, retrieval, monitoring, evaluation, guardrails, lifecycle management, and operational ownership. This broader shift is also reflected in recent systems research, which increasingly focuses on serving frameworks, deployment architectures, and operational efficiency (Majidi et al. 2026)

In practice, the main challenge is therefore not only how to host models locally, but how to provide dependable genAI services under institutional constraints. This connects to middleware-oriented views of LLM deployment, where independent hosting can address privacy, cost, and customization needs, while also creating integration challenges with existing systems (Guran et al. 2024).

Building on these observations, this transition can be understood as a shift across maturity layers, from experimental local prototypes toward institutional services and shared or national infrastructure layers, each with different operational constraints (Table 1).

*Table 1: Maturity layers for institutional LLM hosting*

| Layer | Characteristics | Typical constraints |
|---|---|---|
| Experimental: *local prototypes* | exploratory, flexible | fragile, unsupported |
| Institutional: *supported services* | governed, monitored | Operational expectations (staffing, lifecycle) |
| Shared/national | federated, pooled | coordination, abstraction |

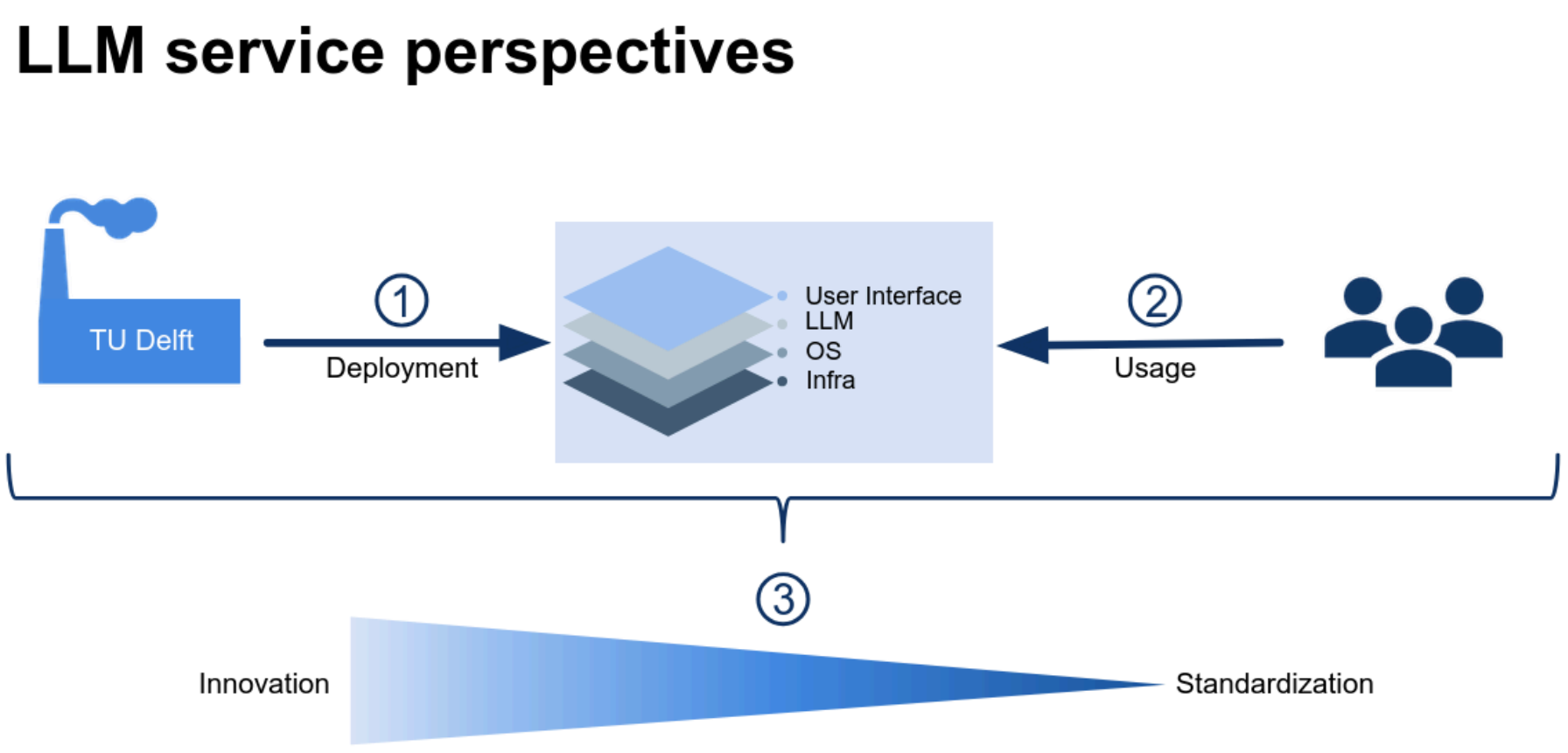

Fig 1: LLM service perspectives - Governance breakout

## 2.3 Emerging ecosystem of institutional LLM hosting

The Dutch academic ecosystem currently includes a growing mix of institutional hosting platforms, national foundation-model initiatives, educational genAI pilots, prompt and workflow libraries, and commercial educational genAI integrations. While some of these initiatives focus on model development rather than hosting, they form part of the broader ecosystem in which academic LLM infrastructure is being shaped.

GPT-NL (https://github.com/GPT-NL) is a national effort to create an LLM from scratch, with a strong focus on open code, open documentation, verifiable or contributed training data, and privacy-preserving data preparation. GPT-NL targets domains including healthcare, education, defense, and the service sector, while positioning the foundational model as a basis for more specialized downstream models. It is expected to be accessible to the Dutch academic community, although not as fully downloadable model-weight distribution through repositories such as Hugging Face.

Similar developments are visible elsewhere in Europe, where several initiatives combine sovereign model development with open science and multilingual capabilities. OpenGPT-X released Teuken-7B (Ali et al. 2025), a multilingual open model trained from the ground up on all 24 official EU languages. The Swiss AI Initiative developed Apertus (Apertus et al. 2025a) as a fully open foundation model for sovereign AI, with open weights, open data, and open science as explicit commitments. TrustLLM (https://trustllm.eu/ ) similarly reflects European interest in trustworthy, open, and sustainable language models for European languages.

Alongside national model-development efforts, the Dutch ecosystem also includes shared infrastructure initiatives such as SURF's AI Hub, formerly WiLLMa, (https://servicedesk.surf.nl/wiki/spaces/WIKI/pages/219086851/AI-Hub+WiLLMa+Pilot+phase ) which aims to provide API-based access to publicly available genAI models in a trusted SURF datacenter environment. At the institutional level, local hosting and genAI service initiatives such as TULIP (https://docs.tulip.tudelft.nl/ ) and Nebula (https://networkinstitute.org/nebula/ ) reflect parallel efforts to develop governed LLM capabilities closer to institutional research and education contexts.

At the European level, initiatives such as Blablador (https://blablador.fz-juelich.de/ ) within the Helmholtz Association illustrate the growing operationalization of publicly governed LLM hosting services. Similar institutional initiatives are emerging elsewhere, including on-premise university platforms, secure self-service environments, institution-specific chat services, knowledge-repository integrations, and domain-specific assistants (Radas et al. 2025; Bhise 2024; Dobur et al. 2024; Hedlund 2024; Schillaci 2024; Basit et al. 2024). Together, these initiatives suggest that academic LLM hosting is evolving from isolated institutional experiments toward a broader ecosystem spanning foundation-model development, institutional hosting, educational integration, commercial access, and shared public-interest services.

Looking ahead, AI factories were also discussed as a possible future layer in the Dutch and European infrastructure landscape. These may provide higher-capability infrastructure and support around requirements and use cases, but their relationship to institutional hosting platforms and shared national services remains to be clarified.

### 2.4 Working definition: LLM hosting as a socio-technical infrastructure

In this paper, we use LLM hosting to refer not only to running inference endpoints, but also to the surrounding infrastructure needed to make these endpoints usable, governed, and sustainable in an academic setting. This includes hardware resources such as compute, storage, and networking; software layers for orchestration, serving, observability, and authentication; and institutional mechanisms for governance, lifecycle management, user-facing abstractions, and operational support. This aligns with broader arguments that access to genAI systems is not only a question of model release, but also of resourcing, technical usability, and utility, and that sovereignty increasingly depends on operational control over the infrastructure needed to deploy and adapt AI systems (Solaiman et al. 2025).

Hosting therefore sits at the intersection of technical systems, governance processes, user support, and organizational capacity. Earlier work on self-hosting in higher education similarly frames digital infrastructure as tied to institutional independence, privacy, data ownership, environmental impact, and resilience, rather than as a purely technical hosting decision (Angeli et al. 2022). In this paper, LLM hosting does not necessarily include foundation-model training, although national model-development initiatives form part of the surrounding ecosystem.

Ethical and governance concerns are often discussed through the lens of privacy, but privacy is only one part of the institutional challenge. Academic LLM hosting also raises questions around access, accountability, transparency, reliability, educational fairness, model lifecycle, and appropriate use.

The relevant distinction is not the form of access or user interface, but who operates the service, how data and access are governed, and what level of support and lifecycle responsibility is provided. Table 2 summarizes three common operating models.

*Table 2: Common operating models for academic LLM hosting and their constraints*

| Operating model | What it offers | Main constraints |
|---|---|---|
| Local / on-premises hosting | Stronger control over data, access, governance, model lifecycle, and institutional service design | Higher upfront cost, staffing, operations, procurement, security, and lifecycle burden |
| External model gateway / commercial API access | Fast onboarding, broad model access, useful for experiments and low-risk use cases | Governance, privacy, budget control, user management, vendor dependency, autonomy, security (export control/dual use), and long-term cost concerns |
| Hybrid / portfolio model | Combines local capability, shared infrastructure, and commercial services based on use case | Requires classification, routing, procurement rules, interoperability, and clear governance boundaries |

# 3 Thematic Insights from practice: State of LLM hosting across Dutch institutions

## 3.1 Infrastructure & Deployment models

Existing academic LLM-hosting initiatives show a range of deployment models, from small-scale institutional platforms to shared national pilots and larger European services. A recurring observation is that many useful serving needs can be met with relatively limited hardware, provided that the service model and expected workloads are clearly understood.

At the individual-use end of the spectrum, tools such as LM Studio (https://lmstudio.ai/ ) or AnythingLLM (https://anythingllm.com/ ) may cover many personal or small-scale use cases, for example on a laptop or workstation. These differ from institutional hosting platforms, where secure access, shared operation, and support for data-sensitive projects or knowledge-management workflows become central requirements.

Nebula at VU is one example of a self-hosted platform for running multiple open-source LLMs locally, motivated by the need to avoid sending sensitive research data to external commercial genAI providers. Staff and researchers can interact with the platform through both an API and a web interface. The hardware setup is modest, with two GPUs and no multi-GPU support yet, but it currently handles the observed load, with a maximum of around 35 concurrent users.

TULIP represents a related institutional hosting effort at TU Delft, running multiple open-source LLMs on a single node with three NVIDIA L40S GPUs. A key difference is its API-first access model, emphasizing programmatic access for institutional research and education use cases.

At the shared infrastructure layer, SURF's AI Hub is being piloted as an LLM-hosting service with API-first access, support for batch jobs, and active/sleep model modes; backed by a slurm-based cluster. This reflects a different deployment model from local institutional platforms, closer to shared national infrastructure for inference.

Blablador provides a larger European example of academic LLM hosting evolving over time. Its working principle was described as starting with what is available and growing gradually. The infrastructure evolved from a small Slurm-based setup with RTX 3090 GPUs, EasyBuild, FastChat, and vLLM, toward use of larger supercomputing and cloud environments, including Kubernetes-based deployment with Synapse and vLLM. Blablador also uses different authentication entry points for industry and academic researchers, limiting which models are available to different user groups.

Across these examples, infrastructure planning is shaped less by user numbers alone than by usage patterns and functionality. Regular end-user activity may be manageable on relatively modest infrastructure, while benchmarking and agent-based use cases can introduce sporadic spikes in resource demand. This has direct implications for capacity planning and cost management. Operational controls such as rate limiting also become important as workloads change. Blablador required rate limiting after agentic workflows became more common, and similar experience was reported for SURF's AI Hub.

These examples also underline that "serving" does not automatically imply 24/7 production availability. Academic deployments may range from experimental endpoints to best-effort institutional services and, eventually, services with stronger uptime, support, and incident-response expectations.

Deployment choices are also constrained by hardware reliability, datacenter limitations, and maintenance practices. GPU reliability was reported to vary across environments and GPU classes; datacenter constraints such as cooling capacity restrict which GPU types can be deployed; and maintenance models such as scheduled shutdowns versus live repair affect service availability. These observations reinforce that LLM hosting is not only a question of selecting models or GPUs, but of matching deployment architecture, workload patterns, and operational constraints.

## 3.2 Energy & Utilization Considerations

Organizations and companies are looking for clearer direction on how to reduce the energy consumption of AI services. For LLM and generative AI hosting, the environmental footprint is closely linked to GPU utilization, power draw, and overall energy consumption, but these are not yet consistently measured or reported.

A recurring gap is the lack of knowledge and integration effort needed to 'glue' existing tools together to capture, report, and interpret LLM-specific operational KPIs. Existing tools, such as Philips' Energy Measurement Tool (EMT) (https://github.com/FairCompute/energy-monitoring-tool), may provide useful starting points for measurement, but institutions still face challenges integrating them into operational LLM-hosting workflows. This makes it difficult to know where energy reduction should be targeted, which operational choices have the largest impact, and when institutions may approach the ceiling of available energy supply.

Several practical issues complicate energy-aware operations. Idle energy waste can occur when demand is low but hardware remains allocated or powered. A wide range of services and hardware types also makes consistent reporting difficult. In addition, siloed hardware across departments or organizations can lead to lower utilization and avoidable energy waste.

Capacity planning is therefore closely connected to usage estimation and admission control. However, long-term service usage remains uncertain, especially as more resource-intensive agentic functions emerge. This makes it difficult to define stable KPIs, estimate future demand, and design energy-aware LLM services.

### 3.3 Governance & Operational sustainability

Governance and operational sustainability emerged as central challenges for institutional LLM hosting. The issue is not only whether institutions can deploy models, but whether they can define rules, responsibilities, support structures, and feedback mechanisms around their use.

A recurring governance problem is the absence of clear guidance. Researchers, students, and teachers are already using LLMs, but institutional rules around what is allowed, what data can be used, and which use cases require additional review are often missing or unclear. This is especially problematic in cases involving research use, student data, or situations where students themselves become research objects in AI-related work.

Some institutions have started to address this through internal AI frameworks developed together with policy teams. For example, TU/e worked on an AI framework with policy colleagues, which helped adoption. However, the broader pattern remains uneven: in many places, “everybody is doing everything,” while guidance, data classification, responsibility boundaries, and research-user policies are still underdeveloped.

Privacy and compliance are among the clearest operational drivers for governed hosting. Commercial API use often requires privacy officers to negotiate contracts, assess data-processing conditions, and monitor compliance. This creates a repeated operational burden when every OpenAI, Google, or similar API key requires separate review. In some cases, data privacy boundaries become the direct trigger for moving away from commercial APIs toward self-hosted or institutionally controlled infrastructure, especially for data-sensitive projects and knowledge-management workflows.

Governance also raises unresolved questions around sovereignty, accountability, and liability. For example, when institutions self-host open models, it may be unclear how responsibility for harmful or dual-use applications should be allocated between model developers, service operators, institutions, or users. Stakeholders may not share a common understanding of what “sovereignty” means in practice, and there is often no education or awareness-building programme around it.

Responsibility also has an organizational dimension. Institutions need to decide what should sit with faculties and research groups, what should be handled by central services, and what requires a shared responsibility model. This is complicated by the fact that teachers and researchers may operate LLM tools in their own environments, creating “black box” usage that is largely invisible to the institution.

User access also depends on objectives, safety concerns, and risk profiles. Students, researchers, teachers, and external partners may use LLMs differently, and the same system may function as a productivity tool in one context and as part of the research setup in another.

Operational sustainability is also a funding and staffing question. Current funding is often tied to specific research contexts, while the potential application of LLM services is much broader; this leaves open the question of how institution-wide LLM use should be structurally funded. Current support roles are often insufficient for genAI and LLM services, and new roles, boards, or committees may be needed as institutions move from experimentation toward more standardized service provision. This includes engineering capacity and knowledge for local services, as well as support capacity for commercial genAI offerings.

Blablador illustrates this operational reality: it began largely as a one-person effort, later expanded to four people, and is hiring additional staff as usage and service expectations grow.

Support organizations are therefore becoming more than infrastructure providers. TU/e’s Supercomputing Center reflects this reality in its support model, positioning itself as a place that helps users choose suitable infrastructure, optimize workloads, and overcome technical or organizational barriers. (TU/e Supercomputing Center: https://supercomputing.tue.nl/). Similar support models may be needed for institutional LLM services, where users need help not only with compute access, but also with model choice, policy interpretation, workload design, and responsible use.

Several governance questions also concern feedback, quality assurance, and lifecycle support. In a portal with multiple deployed models, it remains unclear how institutions should capture feedback on model quality, which metrics should be used, who is responsible for acting on that feedback, and which models should be supported, and for how long.

Overall, governance and operational sustainability require more than policy documents. They require clear roles, data-use rules, support structures, feedback mechanisms, and institutional pathways for moving from innovation to more reliable service provision. Several recommendations emerging from the breakout discussions, including green lanes, risk-based deployment, reuse of existing policies, and agile prioritization of use cases, are therefore treated further in Section 5.

### 3.4 Tooling & ecosystem integration

LLM hosting depends on a wider tooling ecosystem around the model itself, including containerization, serving layers or model runtimes, user interfaces, API gateways, observability tools, and lifecycle management, among others (Figure 2). Retrieval-augmented Generation (RAG) components may also be included. Containerization, for example Docker, is widely used to manage model deployment, especially because different models may require different serving-layer versions or builds, such as different vLLM (Kwon et al. 2023) versions.

Nebula, which was demonstrated during the workshop, illustrates a concrete example of this layered tooling model. It uses Ollama (https://ollama.com/ ) to run models locally, described as "the engine of the car." Once a model is running, different tools can be added around it depending on the user group and service need. Interfaces such as Open WebUI (https://github.com/open-webui/open-webui), LibreChat (https://github.com/danny-avila/LibreChat), Jan (https://github.com/janhq/jan), and LM Studio (https://lmstudio.ai/) serve different types of users. For the chat interface, Nebula uses Open WebUI because it provides granular control over user access, including which users can access which models. Single sign-on is integrated so users can log in with their university accounts.

API gateways and LLM proxies form another important tooling layer. Tools such as LiteLLM (https://www.litellm.ai/) and PortKey (https://portkey.ai/) can provide a unified API interface across many models and providers. Nebula, for example, uses a self-hosted LiteLLM as a proxy or "steering wheel," routing requests to local or external models such as OpenAI, Gemini, Anthropic, and DeepSeek through one endpoint and one API key. This avoids users having to manage many provider APIs separately. The gateway also provides usage information, such as token counts, model utilization, and active users, while intentionally not logging conversation content.

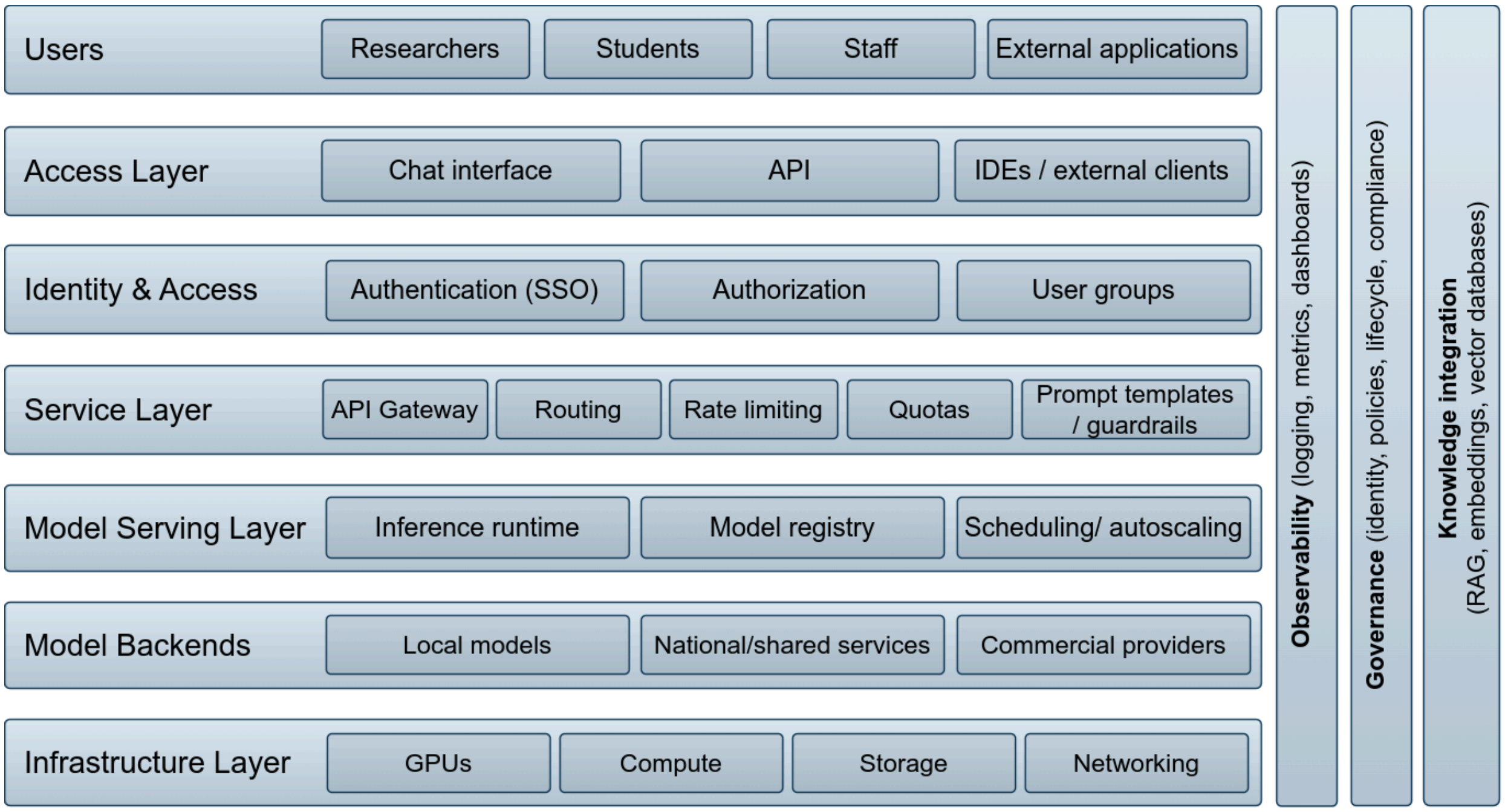

*Fig 2: Representative layered architecture of an institutional LLM hosting platform: Although specific implementations differ, academic deployments typically share the same functional layers, including user access, identity, service orchestration, model serving, infrastructure, and cross-cutting concerns such as observability, governance, and knowledge integration.*

Prompting, routing, and user guidance form another tooling layer. Well-designed system prompts can already support many use cases, but this raises the question of whether users should be expected to write their own prompts. Some institutions address this through templates or prompt libraries, such as Leiden-style templates and the Maastricht University prompt library (https://library.maastrichtuniversity.nl/apps-tools/ai-prompt-library/). Blablador and Nebula use baked-in system prompts for chat interfaces, while leaving API access more open for users with different maturity levels. Agentic environments may also use prompts to generate or improve system prompts, including examples using a “goal achieved” parameter to evaluate task success. These examples reflect a broader tension between flexibility and usability: more powerful systems often require greater user expertise, creating a need for additional guidance, templates, or tooling.

Observability and monitoring tools are also part of the operational stack. cAdvisor, DCGM exporter, and Loki were mentioned as useful services for collecting container, GPU, and log information. Such tools are important because hosting providers need visibility into usage and system behavior without necessarily collecting prompt or conversation content.

RAG is another important, but challenging, integration layer. It can reduce hallucinations by searching through documents before answering, and may be built into interfaces such as Open WebUI or provided as a separate modular component. RAGFlow (https://ragflow.io/) was also mentioned as a RAG solution that may support users in configuring or adapting models for specific document-based workflows. However, RAG as a service introduces several challenges: choosing embedding models, configuring vector databases, extracting document content properly, and giving users insight into the quality of extraction. In one example, the default extractor in Open WebUI was criticized for pulling only headers rather than full document content. LibreChat was also noted as sending files directly to the model without keeping them on the server, leaving a missing step between file upload and text extraction/embedding that may be difficult for users to understand.

The tooling landscape remains fragmented. Some interfaces provide one needed feature but not another; some RAG solutions are too complex, while others are too simple; and different use cases require different combinations of tools. This makes it difficult to identify a single complete solution and often pushes institutions towards building or integrating their own stack. Tooling choices also change over time. Blablador, for example, started with FastChat (https://github.com/lm-sys/fastchat) but is moving away from it because of limited active support, toward a Rust-based vLLM router with more active development and accepted contributions. This illustrates that tooling sustainability and upstream maintenance are part of the operational burden of academic LLM hosting.

## 3.5 Production-Readiness & Commercial Integration

Commercial genAI platforms are already part of the institutional LLM landscape, particularly through educational licensing, public cloud services, managed genAI platforms and initiatives such as EduGenAI (https://npuls.nl/edugenai ), which explore institutionally provisioned access to generative AI tools for education. The key question is therefore how commercial offerings are governed and integrated alongside local and or shared academic capabilities.

Substantial shadow IT already exists: students, researchers, and staff use commercial LLM services such as ChatGPT, Gemini, Grok, and others outside formally provisioned or governed institutional pathways. Making it easier to migrate from personal accounts to institutionally governed access is therefore part of the commercial-integration challenge.

In this context, commercial solutions can serve as accelerators for production-readiness. NVIDIA AI Enterprise and NIM offer one pattern, close to institutions that already invest in NVIDIA-based infrastructure, with production-oriented inference and model customization workflows (https://docs.nvidia.com/ai-enterprise/index.html ). Microsoft Azure AI Foundry represents another pattern, providing a broad platform for building, operating, governing, monitoring, and evaluating genAI applications and agents (https://learn.microsoft.com/en-us/azure/foundry/what-is-foundry).

For Dutch higher education, these commercial options arrive in an ecosystem where SURF already supports institutions through public-cloud procurement, Microsoft-focused innovation and knowledge-sharing programmes, and guidance on cloud-agnostic, cost-effective, and secure use of public cloud services in research settings. SURF's emphasis on public values, including transparency, also illustrates that commercial integration is not only a technical or procurement question. The strategic question is how to adopt commercial platforms without undermining interoperability, procurement flexibility, or longer-term digital sovereignty (https://www.surf.nl/en/themes/procurement-contracting/microsoft-inspiration-and-innovation-programme-miip).

Commercial integration also introduces several practical gaps. These include hardware vendor lock-in, limited engineering capacity and knowledge, support requirements for commercial offerings, security and privacy regulations around sensitive data and GDPR, AI use-case classification procedures for routing users to local or commercial offerings, and procurement processes for commercial genAI services.

Production-readiness therefore depends not only on choosing a commercial or local platform, but on defining how different offerings are governed, procured, supported, monitored, and matched to institutional use cases.

## 3.6 Research & Education Use Cases

Research and education use cases show why academic LLM hosting cannot be designed around a single generic access pattern.

Education use cases already include personalized learning, exercise generation, Socratic tutoring, and course-level support. Examples include pilots for Socratic AI tutors, collections of educational genAI pilots (e.g., UvA pilots: https://tlc.uva.nl/en/article/genai-roles-in-your-course/), and prompt libraries (e.g., Maastricht library https://library.maastrichtuniversity.nl/apps-tools/ai-prompt-library/ ) intended to help teachers and students use generative AI more effectively.

In teaching contexts, one recurring requirement is that access should be reliable, fair, and aligned with course design. Personalized learning use cases may require LLMs that take students' knowledge and skills background into account, while exercise-generation use cases require predictable access during teaching periods. Keeping educational material protected from exposure to commercial LLM providers was also raised as an important concern. Education use also raises questions about trust and assessment, including whether resource-intensive formats such as handwritten, oral, or in-person exams are needed in some contexts.

Research use cases are similarly diverse. LLMs are used for coding support, natural language processing (NLP)-oriented workflows, knowledge management, document processing, and other research tasks. At the same time, LLMs may themselves be the subject of research, requiring experimentation, evaluation, customization, and domain-specific adaptation. There appears to be substantial use of commercial LLM services already, but the level and nature of this use are not always visible institutionally.

Research use also introduces specific governance and infrastructure requirements. AI policies in research contexts are not always clear, and secure hosting is especially important for data-sensitive projects and not-yet-open or non-public research where keeping data confidential is a high priority.

These use cases point to a broader infrastructure requirement: academic LLM services need to support different users, risk profiles, and access modes across education and research, rather than assuming one model, interface, or governance pathway will fit all contexts.

# 4 Cross-cutting tensions and systemic patterns

## 4.1 Control vs Operational Burden

A recurring tension across the workshop is that greater institutional control over LLM services also increases operational responsibility. Local or institutionally governed hosting may improve control over data, models, access, and compliance, but it also requires funding, staffing, monitoring, user support, and long-term service ownership.

Funding is one example of this tension. Current funding is often limited to specific research contexts or domains, while the potential application of LLM services is much broader (problem/opportunity). This creates a gap between project-based funding and institution-wide demand. Rather than being a binary blocker, funding often determines the feasible service scope, reliability expectations, and sustainability model, especially where there is no structural funding model for broader LLM use.

Transparency and privacy create a second tension. Operational visibility is needed to improve services, monitor usage, understand performance, and report energy or utilization metrics. At the same time, some forms of operational logging, particularly those linking user identities to prompts, conversations, or detailed usage patterns, may create additional privacy, GDPR, and AI Act considerations. The more detailed the operational telemetry, the easier it becomes to optimize the service; the more it reveals about individual users or their activities, the greater the governance burden becomes.

Sovereignty also introduces operational questions. Community-owned or institutionally governed tools may advance digital sovereignty, but sovereignty is not achieved by deployment alone. Stakeholders may have different understandings of what sovereignty means, and low awareness can lead to under-use, misuse, or unrealistic expectations of local hosting.

Access control creates a related trade-off. Separating user groups can improve security, compliance, and model governance, especially when different users should have access to different models or capabilities. However, such separation can create parallel access paths, tooling, and operational processes ecosystem, increase integration costs, and make the service harder to operate consistently.

Model choice and service scope add another layer of burden. Institutions may receive requests from domains such as medicine, where a general model may not be suitable for specialized documents or sensitive workflows. Supporting such use cases increases the value of institutional control, but also raises questions about which models to host, how far to specialize, and who is responsible for evaluating fitness for use.

Finally, there is a tension between powerful tooling and ease of use. Many LLM users may not know how to define good system prompts or configure parameters such as temperature and context length. Some organizations address this through templates, while systems such as Blablador and Nebula use built-in system prompts for chat interfaces while leaving API access more open for advanced users. Agentic environments may also use prompts to generate or improve system prompts dynamically. These examples show that control and flexibility often require additional user guidance, defaults, and tooling if services are to remain usable for non-expert users.

Overall, institutional control is not only a technical capability. It creates continuing obligations around funding, privacy, monitoring, model selection, user segmentation, support, and guidance.

### 4.2 Flexibility vs production readiness

A second recurring tension concerns the movement from innovation toward standardization. LLM services are still evolving quickly, and it is not always clear which problem institutions are trying to solve, which use cases should be prioritized, or whether standardization is even achievable or desirable given cost and complexity.

Standardization may help institutions provide more reliable services, clearer governance, and more predictable support. At the same time, over-standardization can stifle innovation, especially when LLM use cases differ substantially across research, education, and support contexts. Whether it is desirable to move from innovation to standardization therefore depends on the use case; different use cases may require different standards.

This tension also appears in the distinction between models and platforms. Institutions must decide whether the priority is to provide access to particular models, to build stable platforms around them, or to support a broader LLM service ecosystem spanning chat and API access, local and commercial tools, integrations, licensing arrangements, and supporting infrastructure. and access arrangements (Figure 3). As the number of users grows, so do the use cases. No single interface or stack covers all requirements: some interfaces provide one feature but not another; some RAG solutions are too complex, while others are too simple. This creates pressure to build or integrate local stacks, while still maintaining enough stability for production use.

Model selection is another example of the same tension. A recurring theme was the importance of choosing the right model for the task rather than defaulting to the largest available option. Mixture-of-Experts models and smaller models were discussed in this context, with the observation that “bigger is not always better.” Smaller or faster models may provide sufficient performance at lower cost, while larger models may be needed only for specific tasks.

This raises questions about defaults and user guidance. One suggestion was to make smaller models the default and allow users to opt into larger models when needed. Such an approach could preserve flexibility while helping manage costs and capacity. It also raises a responsibility question: should platforms such as the AI Hub actively educate users on model selection, especially when users may not naturally consider cost-performance trade-offs?

Production readiness also depends on lifecycle and model-governance decisions. New models arrive constantly, and keeping up with the state of the art requires continuous monitoring of model releases, benchmarks, licenses, and community signals. Maintaining a diverse set of models (e.g., small, fast, large, specialized, or general) increases flexibility, but also creates hard choices about what not to host. Users may need to accept that not every model can be made available through an institutional service.

Legal, licensing, and value considerations further constrain flexibility. Not every technically capable model is suitable for institutional deployment. Model provenance, licensing terms, jurisdiction, transparency, bias, and alignment with organizational values may all affect which models can be offered. For example, SURF’s emphasis on public values, including transparency, illustrates how production-readiness is shaped not only by performance and cost, but also by institutional principles.

Overall, the challenge is not simply to standardize LLM hosting, but to decide where standardization is useful, where flexibility must remain, and how institutions can support changing use cases without turning every deployment into a bespoke project.

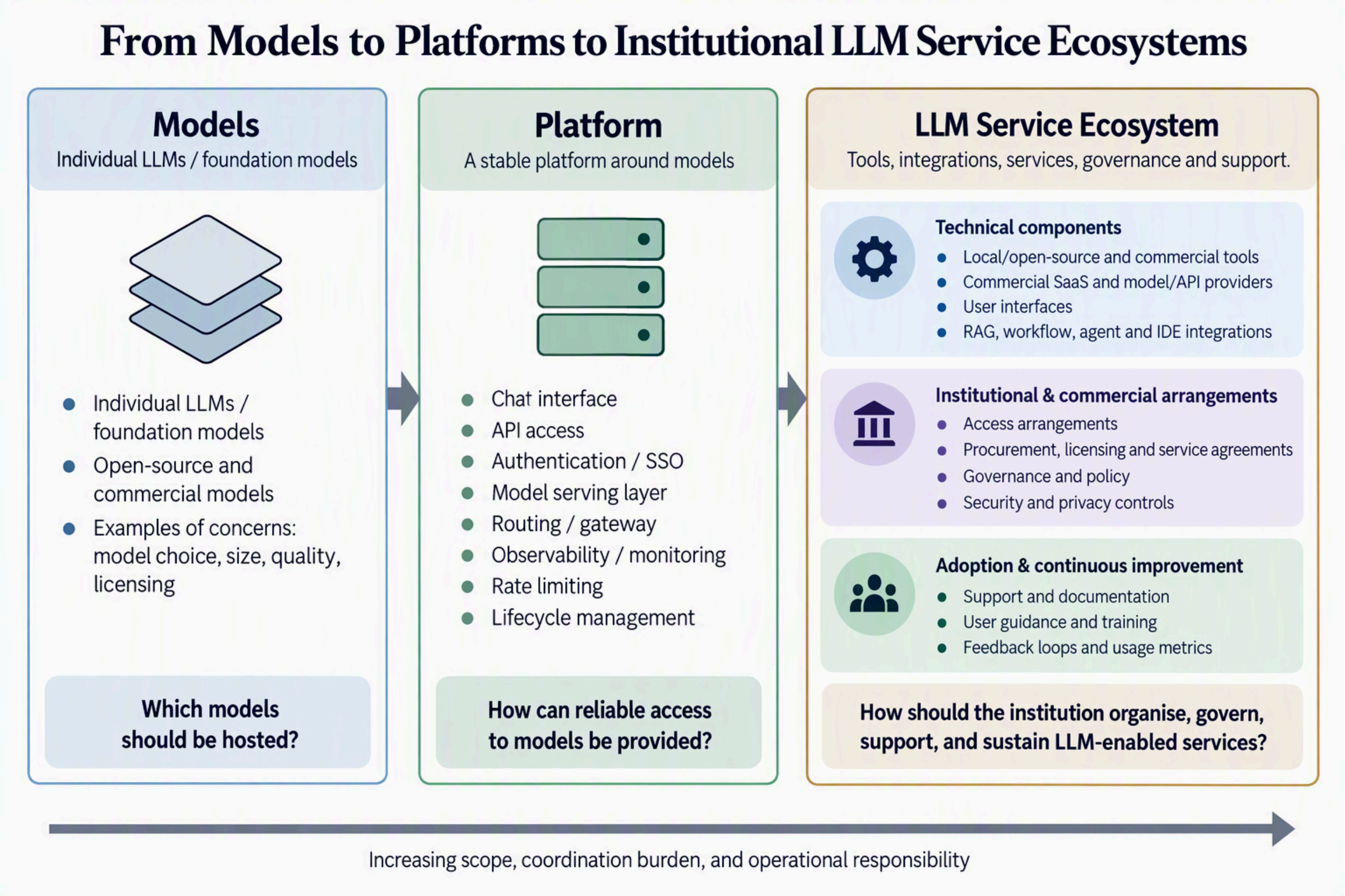

*Fig 3 From models to platforms to institutional LLM service ecosystems. Visual generated with Microsoft Copilot based on an author-defined conceptual structure and reviewed by the authors.*

### 4.3 Open ecosystem vs commercial maturity

A third cross-cutting tension concerns the balance between open academic ecosystems and the maturity of commercial genAI offerings. Commercial platforms can accelerate adoption and provide access to high-performing models, managed tooling, and production-oriented services. At the same time, their use raises questions around privacy, security, sovereignty, cost, and institutional dependence.

This tension is especially visible in the gap between open-source and proprietary model capabilities. The gap is real, but several discussions suggested that the institutional goal does not always need to be matching frontier commercial models. Being “six months behind” with models that are reliable, safe, cheaper, or easier to govern may be acceptable for many academic use cases. This creates a trade-off between using the newest proprietary models, with higher performance and often higher cost, and using open-source alternatives that may lag behind but offer more control.

Another dimension of contrast is service expectations. Academic data centers and HPC environments may rely on maintenance practices such as scheduled shutdowns rather than live repairs, while commercial platforms often offer stronger availability and support expectations through managed-service models or SLAs.

Commercial use in academic settings also raises questions about scholarly independence and exploitation. Allowing commercial integration can bring funding, industry relevance, and access to mature services, but it may also threaten the openness and independence of academic work. Research and academic use of commercial systems therefore cannot be treated only as a procurement or performance question; it also involves academic values, data governance, and long-term dependency.

Institutions also need support capacity for commercial offerings. Even when a commercial platform is mature, institutions may still lack the internal support structures needed to help users adopt it safely and effectively.

The same tension appears in education. There is a debate about whether students should be taught with cutting-edge tools or deliberately limited ones. Access to frontier tools may better reflect industry practice, but understanding the weaknesses of limited or open models can itself be a learning objective. This leaves open the broader question of what responsibility universities have in preparing students for a world where industry may use more powerful tools than academia can provide.

Overall, the question is not whether open or commercial systems are categorically preferable. The more practical tension is how to balance innovation acceleration through commercial offerings with privacy, security, pedagogical, and sovereignty requirements. This balance will likely differ by use case, risk profile, and institutional responsibility.

## 4.4 Institutional capability vs shared infrastructure

A fourth cross-cutting tension concerns what should be hosted locally, what should be provided by central institutional services, and what should be shared at national or cross-institutional level. This is both a technical and governance question: faculties, central services, and national infrastructure providers may each be appropriate for different parts of the LLM hosting ecosystem.

Shared infrastructure can improve efficiency compared with fixed, siloed hardware allocations, both from an energy perspective and from an operational perspective. Centralized services such as SURF's AI Hub may be well placed to provide access to large models, shared fine-tuning capabilities, or other services that benefit from pooled capacity and expertise. At the same time, community-developed or jointly operated platforms across institutions may offer a complementary path, particularly where flexibility, experimentation, and shared ownership are important.

At the same time, domain-specific models complicate purely centralized approaches. Domain-specific fine-tuned models may provide a strong business case for centralized LLM hosting, especially where several institutions share similar needs. However, some models may be too specific, experimental, or locally governed to make sense as national services. In such cases, institutions may still need their own platforms for smaller, specialized, or sensitive deployments.

One possible coordination model is for the academic community to decide collaboratively which organization hosts which models. This could reduce duplication and support a more distributed model ecosystem. However, reaching agreement on such coordination is difficult in practice, especially when institutions have different use cases, risk profiles, funding models, uptime expectations, and operational responsibilities.

Routing strategies similar to OpenRouter ("OpenRouter," n.d.) were also discussed, where requests could be dynamically directed to local infrastructure or centralized services such as SURF. This also introduces challenges around uptime guarantees and operational responsibility, especially when relying on external or distributed infrastructure.

A hybrid ecosystem may therefore be more realistic than a single centralized approach to providing genAI services. In such an ecosystem, national or shared services provide access to larger models or common capabilities, while local deployments handle smaller, specialized, or institution-specific models. Examples such as PublicAI at ETH Zurich (Apertus et al. 2025b) were mentioned in this context and may provide useful comparators.

Overall, the question is not whether LLM hosting should be local or centralized, but how institutional capability and shared infrastructure can be combined without creating unclear ownership, duplicating effort, or weakening governance.

### 4.5 Performance ambition vs sustainability constraints

A fifth cross-cutting tension concerns the gap between expanding genAI capability needs and the limits of local infrastructure, engineering capacity, software maturity, and energy availability. Institutions are asked to support increasingly capable LLM services, but production readiness depends on more than access to GPUs.

Hardware acquisition is a major constraint. Blablador's experience highlighted hardware as one of the most expensive parts of building LLM infrastructure, alongside constraints around GPU vendors and hardware availability. The cost of obtaining sufficient accelerator memory and compute capacity can itself limit which models and workloads institutions are able to support.

Reliability also varies across environments and GPU classes: some deployments reported failures with GPUs such as L40S and RTX4000, while Blablador noted that consumer-grade GPUs can be surprisingly stable and that some high-end GPUs, such as Grace Hopper, showed frequent failures.

From an energy perspective, the main issue is not only hardware efficiency, but whether capacity is actually used well. Environmental footprint depends on GPU utilization, power consumption, and energy use, yet energy-related measurements remain difficult to integrate into systematic reporting pipelines for data center engineers.

A related cost nuance is that energy prices can change the economics of local hosting. One concern raised in subsequent workshop discussion was that, in the Dutch context, high electricity prices may make token-based commercial access cheaper than local hosting for some low-utilization or intermittent workloads.

Availability and lifecycle expectations add further pressure. Once LLM services become embedded in research and education workflows, users may come to expect stable access, limited downtime, and continuity across model versions, even when the service was originally closer to an experimental endpoint or best-effort institutional service. Maintaining such continuity can conflict with hardware capacity and maintenance, software upgrades, model deprecations, and the pressure to keep up with newer models.

Scaling challenges are not driven only by the number of users. Experience from AI Hub, corroborated by Blablador, suggests that functionality and usage patterns matter strongly. Benchmarking and agent-based workloads were described as highly resource-intensive and "spiky," often dwarfing regular end-user activity. Long-term service usage is therefore also difficult to predict, especially as agentic functions become more common. This has direct implications for infrastructure planning and cost management.

User expectations can also inflate infrastructure demand. Some users may expect LLM infrastructure to solve tasks that are better handled by more specific or traditional methods, such as OCR or custom computer-vision models for visual document processing. Similarly, demand for domain-specific fine-tuning, for example in healthcare or cancer-treatment contexts, is seen as desirable but complex, especially when combined with RAG or sensitive-data requirements.

Overall, the tension is that institutions want to support more capable, specialized, and interactive AI workloads, while infrastructure planning must account for energy limits, hardware cost, reliability, workload spikes, uncertain future usage, availability expectations, model lifecycle, and limited engineering capacity.

## 5 Emerging directions & coordination opportunities

The workshop discussions point to several emerging directions rather than a single preferred model for academic LLM hosting.

1. Maintain focus on inference and service capability:
   Academic institutions do not need to compete with commercial services, nor aim to become a European equivalent of Hugging Face. The more immediate focus is model inference, institutional service provision, and the capability to use, govern, and support LLMs effectively.

2. Prioritize skills, usable defaults, and feedback loops:
   Institutions should balance keeping up with relevant model releases against investments in user skills, guidance, and usable defaults. This includes accessible training, prompt templates, default model configurations, guidance toward smaller models where appropriate, and different defaults for chat vs API users. Portals with multiple deployed models should also include mechanisms for collecting feedback on model quality, usage, and user needs through surveys, user panels, usage analytics, or other feedback channels.

3. Classify use cases and develop risk-based service pathways:
   Institutions need procedures to classify genAI use cases by data sensitivity, reproducibility needs, architecture, user group, and risk profile. This can help route users toward local hosting, shared infrastructure, or commercial offerings. Low-risk or well-understood uses could follow pre-approved “green-lane” routes, while sensitive or high-impact uses would require deeper review. These pathways could build on existing trusted services and should involve Authorised Person (AP) or Data Protection Officer (DPO) expertise early for data-sensitive research and education use cases.

4. Reuse and adapt existing policies where possible:
   Rather than creating entirely new processes for every AI use case, institutions can build on existing privacy, security, research-data, and procurement processes. At the same time, clear AI policies are still needed at national, inter-institutional or regional, and institution-specific levels, alongside secure hosting setups for data-sensitive projects and knowledge-management use cases.

5. Govern commercial adoption as part of a portfolio:
   Commercial offerings can reduce setup time, improve operational maturity, and provide monitoring or evaluation capabilities that institutions would otherwise need to assemble themselves. At the same time, they introduce dependencies on particular ecosystems, interfaces, and infrastructure assumptions. Academic institutions should therefore govern commercial adoption centrally, classify relevant use cases explicitly, and design services around portable interfaces and exit options wherever possible. Commercial adoption pathways should also reduce shadow IT by making it easier for users to migrate from personal accounts to institutionally governed services.

6. Mature operational service practices:
   Once an LLM service exists, institutions need documentation, operational procedures, user support, and ongoing digital/AI skills development. Alias-based access can reduce dependency on hardcoded model names and make model lifecycle changes easier. Rate limiting may be required as agentic workflows increase load. Model portfolios should be deliberate: for example, offering one strong general model and one smaller high-throughput model, while encouraging users to choose the smallest model that meets their needs. At the same time, researchers and other institutional users should have access to high-capability models where needed for coding, knowledge management, agentic tool use, or other demanding workflows. A board or similar governance body may be needed to support the transition from innovation to standardization, especially for use cases with reputational or institutional risk. Users also need to be informed about practical service constraints, including how long models can be supported, what changes over time, and what lifecycle guarantees are realistic.

7. Coordinate community-owned and community-governed infrastructure:
   GPT-NL's content board ( https://gpt-nl.nl/samenwerken/content-board/ ) provides one example of coordinating data contributors and governance around data collection. More broadly, community-owned or community-governed genAI platforms may strengthen digital sovereignty, but this requires awareness-building, outreach, and adoption support. Outreach also matters for service uptake: examples such as talks, internal communication, conferences, and informal community-building were noted as important for making services visible and usable.

8. Align services with organizational values and transparent metrics:
   SURF's emphasis on public values, including transparency, illustrates the importance of designing genAI services around more than performance alone. Transparency also applies operationally: institutions may need metrics on unique users, request counts, costs, and service usage to convince management and improve operations, while avoiding unnecessary logging of prompt content or privacy-sensitive data.
   The metrics supplied by tools also need explanation, so that usage, cost, and performance data can be interpreted correctly by managers, service teams, and users.

Overall, the emerging direction is a portfolio of service pathways: local and open capabilities where control, privacy, or reproducibility matter; shared or national services where pooled capacity is useful; and commercial platforms where they clearly improve time-to-value or service quality. The coordination challenge is to make these pathways legible, governed, and usable without forcing every institution or research group to solve the same problems independently.

## 6 Open questions and unresolved tensions

Several open questions remain unresolved. These are not only technical questions, but also governance, educational, legal, and institutional questions that require further discussion across infrastructure providers, researchers, teachers, students, policy teams, and national actors.

1. What role should LLMs play within universities?
   A basic ethical question remains: should LLMs be used within universities, and if so, under which conditions? Some groups may prefer to wait until systems are considered sufficiently safe or ethically mature, while others are already embedding them in research, education, and institutional workflows. This creates a gap between actual use and institutional consensus.

2. How should education and assessment adapt?
   Educational use raises unresolved questions around trust, fairness, and assessment design. Suggested directions included more verbal, handwritten, oral, or in-person assessment formats, as well as summative assessments that evaluate the student's conversation with genAI rather than only the final output. However, these approaches can be resource-intensive and require further pedagogical discussion. A broader question remains: what responsibility do universities have in preparing students for a world where industry may use more powerful tools than academia can provide?

3. How should institutions handle logging, feedback, and consent?
   Useful feedback, usage data, or interaction traces may help improve services, education, and research. However, logging is sensitive, especially in educational and research contexts. Open questions include whether users can donate logs, whether logging should be constrained by place or time, whether users should be able to remove sensitive parts before sharing, and how summaries can be used without violating privacy, trust, or power relations.

4. What does responsible governance require beyond privacy?
   Ethics is often framed primarily through privacy, but privacy is only one part of the challenge. Institutions still need clearer guidance on what is allowed, what data may be used, which use cases require additional review, and how responsibilities are distributed. Ambiguous or missing policy creates legal and ethical risk for hosting. The EU AI Act is relevant, but its operational implications for academic service providers remain unclear, and institutional genAI policies are still developing.

5. What does digital sovereignty and digital autonomy mean in practice?
   Limited awareness of digital sovereignty and autonomy among stakeholders remains an issue. Community-owned or community-governed platforms may strengthen sovereignty, but only if users and decision-makers understand what sovereignty means operationally. Related openness questions remain, for example when a model may be accessible to the academic community but not downloadable in the same way as models distributed through repositories such as Hugging Face.

6. How should institutions govern the transition from pilots to standardized services?
   Institutions still need governance processes for moving LLM services from pilot use cases toward wider adoption. Open questions include how to deploy institutional LLM services securely, scalably, and compliantly; how to ensure responsible use aligned with legal and institutional policies; and how to decide when a service is ready to move from innovation to standardization.

7. How should users be guided toward the right model or service pathway?
   Users need help selecting appropriate models without being overwhelmed. Open questions include how much responsibility service providers should take for user education, how model choice should account for cost, performance, energy, and risk, and how to route users between local hosting, shared infrastructure, and commercial offerings.

8. How should infrastructure handle agentic, multimodal, and specialized workloads?
   The rise of agentic AI creates new guidance needs for autonomous or semi-autonomous systems. Multimodal serving also remains underdeveloped: image-generation and visual-document workflows may expose APIs similar to text models, but behave very differently operationally. Some use cases may be better served by traditional OCR or custom computer-vision models rather than LLM-augmented pipelines. Domain-specific fine-tuning, for example in healthcare or knowledge-management contexts, is desirable but complex.

9. Can a federated model ecosystem realistically be coordinated?
   A federated ecosystem could allow different institutions or national services to host different models or capabilities. However, it remains unclear whether such coordination can work in practice, especially given differences in uptime expectations, operational responsibility, funding models, data sensitivity, and local priorities.

10. What is the future role of national and European infrastructure?
    Future national and European initiatives, including AI factories, may provide higher-capability infrastructure, support, and collaboration opportunities. However, their relationship to institutional hosting, local support teams, and shared service pathways remains unresolved. Further work is needed to clarify what should be provided locally, institutionally, nationally, or through European infrastructure.

11. How should institutions balance local control with commercial capability?
    Not every academic use case requires local or open LLM deployment. Commercial production-ready genAI solutions may be appropriate in some cases, provided that institutions maintain a critical stance toward vendor lock-in, FAIR principles, privacy, security, and digital sovereignty. The unresolved question is how to balance innovation acceleration through commercial offerings with long-term institutional control and public values.

# 7 Conclusions

Academic LLM hosting is an emerging socio-technical infrastructure question for universities, not only a matter of model access or experimentation. The workshop discussions showed growing demand across research, education, and institutional support, alongside recurring challenges around infrastructure and deployment models, energy and utilization, governance and operational sustainability, tooling and ecosystem integration, production-readiness and commercial integration, and research and education use cases.

A central lesson is that LLM hosting cannot be treated as a single technical deployment model. Institutions face choices between local hosting, shared or national infrastructure, commercial access, and hybrid combinations of these service pathways. Each option involves trade-offs between control and operational burden, flexibility and production readiness, open ecosystems and commercial maturity, institutional capability and shared infrastructure, and performance ambition and sustainability constraints.

The relevant question is therefore not whether academic institutions should choose one model over all others, but how they can build governed, usable, and sustainable service pathways for different needs and risk profiles.

Consequently, the emerging directions are practical rather than uniform: maintaining focus on inference and service capability; prioritizing skills, guidance, and usable defaults; classifying genAI use cases and mapping them to service pathways; developing risk-based and green-lane routes; reusing existing policies; building user guidance and feedback loops; governing commercial adoption as part of a portfolio; maturing operational service practices; coordinating community-governed infrastructure; and aligning services with organizational values and transparent metrics. At the same time, several questions remain unresolved, including the role of LLMs in universities, how education and assessment should adapt, how to handle logging and consent, what governance requires beyond privacy, how to support agentic and multimodal workloads, how to coordinate federated model ecosystems, and how to balance local control with commercial capability.

## 8 Acknowledgements

This work was supported by 4TU.NIRICT. We thank the members of the *Towards Democratized Academic LLM Hosting Workshop* Consortium and the wider workshop community for their contributions to the discussions, breakout sessions, and panel.

***Towards Democratized Academic LLM Hosting Workshop Consortium:** Nick Brummans (TU/e), Paul Jansen (SURF), Aurin Vos (WUR), Radu Apsan (VU), Arezoo Alizadeh (TU/e), Wouters Ras (TU Delft), Bertine van Deyzen (EduGenAI), Duncan Kampert (SURF), Lolke Boonstra (TU Delft), Pamela Wochner (TU Delft), Santosh Ilamparuthi (TU Delft), Niels Taatgen (RUG), Arie van Deursen (TU Delft), Damian Podareanu (SURF), Corry Wouters (TU/e), Thijs van der Plas (WUR).